\documentclass{article} 
\usepackage{amssymb}
\usepackage{amsmath}
\usepackage{latexsym}
\usepackage{verbatim}
\usepackage[english]{babel}
\usepackage{graphicx}
\usepackage{esvect}
\usepackage{cite}
\usepackage{amscd}
\usepackage{epsfig}
\usepackage{pstricks,pst-node}
\usepackage[margin=1in]{geometry}

\def\Journal#1#2#3#4{{#1} {\bf #2}, #3 (#4)}

\def\PRL{\em Phys. Rev. Lett.}

\def\be{\begin{equation}}
\def\ee{\end{equation}}
\def\bea{\begin{eqnarray}}
\def\eea{\end{eqnarray}}

\begin{document}
\title{Measuring Relativistic Dragging with Quantum Interference}

\author{Y. Bonder and J. E. Herrera-Flores}
\date{Instituto de Ciencias Nucleares, Universidad Nacional Aut\'onoma de M\'exico\\
Apartado Postal 70-543, Cd. Mx., 04510, M\'exico}

\maketitle
\begin{abstract}An experiment to test for relativistic frame dragging effects with quantum interferometry is proposed. The idea that the classical trajectories of the interferometer surround a spherical mass source whose angular momentum is perpendicular to the plane containing these trajectories. A simple analysis allows one to find the phase shift for particles traveling in the innermost stable circular orbit; the result can be easily generalized for more realistic orbits. The phase difference goes like the source's angular momentum per mass times the quantum particles' mass. This is a small effect but it can be amplified by making the classical paths go around the source several times. Moreover, this experiment has the advantage that the source's angular momentum can be easily controlled.
\end{abstract}

General Relativity is the accepted theory of gravity and its predictions have been empirically verified in many situations \cite{Will}. On the other hand, Quantum Mechanics is the current theory of matter and the nongravitational interactions. However, these theories are incompatible and the quest to combine them into a theory of Quantum Gravity (QG) is still ongoing. Moreover, it is widely believed that the regime where the QG effects are expected to dominate is at Planck scale. However, this scale is hard to access experimentally.

Perhaps one of the issues delaying the construction of a universally accepted QG theory is the lack of experimental guidance. Still, it is possible to get relevant clues of QG by testing gravity with quantum systems, i.e., systems that cannot be described classically. The first experiment of this kind is the famous COW experiment where the interference pattern of neutrons is affected by gravity \cite{COW}. However, this experiment is insensitive to spacetime curvature: the ``gravitational'' effect measured in the experiment is related to the fact that noninertial frames are used, and tidal effects, which are directly related with curvature, are negligible. This becomes evident by the fact that the same interference pattern is obtain by accelerating the interferometer \cite{NoninertialCOW}. It should be emphasized that, in the modern understanding of gravity, only curvature effects are considered to be gravitational.

Atom interferometry has become an extremely sensitive technique that has led to exquisite measurement of fundamental constants  \cite{Bouchendira} and to stringent bounds on dark energy \cite{Muller}. In addition, it has provided with very sensitive tests of gravity. In fact, such experiments are sensitive enough to measure gravity gradients \cite{Asenbaum}, which are generated by a nontrivial spacetime curvature. In this work a new experiment is proposed that could be sensitive to curvature effects with quantum systems. The proposal is to look for frame dragging effects using quantum interferometry and it is described in the remaining of this work. The notation and conventions of Ref. \cite{Wald} are used.

Consider a spacetime whose geometry is sourced by a spherical and slowly rotating mass distribution. Then, to first order in the angular momentum $J$, the spacetime metric outside this source, in conventional Schwarzschild coordinates $(t,r,\theta,\phi)$, and using natural units, takes the form
\begin{equation}
 d s^2= -\left(1-\frac{2M}{r}\right) d t^2+\left(1-\frac{2M}{r}\right)^{-1} d r^2+r^2 d \theta^2+r^2\sin^2\theta d \phi^2-\frac{4J}{r}\sin^2\theta d t d \phi,
\end{equation}
where $M \geq 0$ is the Schwarzschild parameter which, in this case, is essentially the mass of the spherical source, and the angular momentum points in the direction of $\left(\partial/\partial \phi\right)^a$. Clearly, the source density is such that $r=2M $ is inside it, which is a region that is not covered by the coordinates. Also, note that one recovers the Schwarzschild metric when $J=0$.

In the WKB approximation, the particles' paths in the interferometer are approximated by the classical trajectories \cite{WernerBook}. Moreover, in the test particle approximation, this trajectories are given by timelike geodesics. Solving the geodesic equation is generically hard, however, there are well-known methods \cite{Hartle} to simplify these equations when symmetries are present.

 The spacetime under consideration is stationary and axially symmetric, therefore, $\xi^a=\left(\partial/\partial t\right)^a$, and $\psi^a=\left(\partial/\partial \phi\right)^a$ are Killing vector fields, for which there are two conserved quantities along the geodesics
\begin{eqnarray}
\label{eConserv}
e&=& -g_{ab}\xi^a u^b = \left(1-\frac{2M}{r}\right)\dot{t}+\frac{2J}{r}\sin^2\theta \dot{\phi},\\
\label{lConserv}
l&=& g_{ab}\psi^a u^b = -\frac{2J}{r}\sin^2\theta \dot{t}+r^2\sin^2\theta \dot{ \phi},
\end{eqnarray}
where $u^a$ is the four-velocity with respect to proper time $\tau$ and the overdot represents the derivative with respect to $\tau$. Moreover, this spacetime metric is invariant under $\theta\rightarrow \pi-\theta$, thus, any geodesic that travels on the $\theta=\pi/2$ ``equatorial'' plane for an open time interval, must remain in that plane indefinitibly.

For timelike equatorial geodesics, the four-velocity normalization condition can be casted into the form
\begin{equation}\label{eff eq}
\varepsilon = \frac{1}{2}\dot{r}^2+ V_{\rm {eff}},
\end{equation}
with $\varepsilon = (e^2-1)/2$ and
\begin{equation}\label{eff pot}
V_{\rm {eff}} = -\frac{M}{r}+\frac{l^2}{2r^2}-\frac{Ml^2}{r^3}+\frac{2elJ}{r^3}.
\end{equation}
Thus, the problem of solving the geodesic equations has been reduced to an effective one-dimensional problem described by Eqs.~(\ref{eff eq}) and (\ref{eff pot}). It should be mentioned that the key step in reducing the problem is to invert Eqs.~(\ref{eConserv}) and (\ref{lConserv}) and write $\dot{t}$ and $\dot{\phi}$ in terms of $r$ and the conserved quantities.
 
For simplicity, only an innermost stable circular orbit (ISCO) is considered; the analysis presented here can be easily generalized for more realistic trajectories. However, such generalizations may require numerical methods. The condition for an ISCO is that $V_{\rm {eff}}$ has one extremum, which implies $l=\sqrt{12}M- eJ/M$. Moreover, the ISCO coincides with this extremum, therefore the (constant) radius of the geodesic is $r_{\rm ISCO}= 6M - \sqrt{12} eJ/M$. The proper time along an equatorial ISCO segment is given by
\begin{equation}
\tau = \int \sqrt{\left(1-\frac{2M}{r_{\rm ISCO}}\right) d t^2-r^2_{\rm ISCO} d \phi^2\pm \frac{4J}{r_{\rm ISCO}} d t d \phi }\ ,
\end{equation}
where the integral is along the path and the sign in the last term is $+$ when the trajectory is counterclockwise, as seeing from the $0< \theta <\pi/2 $ region, and the oposite sing when traveling clockwise. This sign change is a consequence of the well-known relativistic dragging that has been measured by Gravity Probe B \cite{GPB}, and, importantly, it is a curvature effect. In other words, there is a proper time difference between equatorial ISCO segments if they are travelled counterclockwise or clockwise.

For a quantum interference experiment where the two classical paths under consideration that go from $\phi=0$ to $\phi=\pi$: one counterclockwise and the other clockwise. The proper time difference between these paths, to first order in $J$, is then
\begin{equation}\label{20}
\Delta\tau=\frac{2\sqrt{2} \pi}{3}\frac{J}{M}.
\end{equation}
If $m$ is the mass of the quantum particle then, the phase difference becomes \cite{WernerBook} $\Delta\varphi = m \Delta\tau$. Clearly, $\Delta\varphi=0$ when $J=0$, as expected.

Of course, in realistic experiments $J$ is expected to be very small, thus, the only hope to detect the predicted phase difference is that the classical paths travel through the region generating the phase difference many times, assuming that quantum coherence can be maintained. Still, a nice feature of the proposed experiment is that $J$ can be easily modulated to help extract the signal.

Finally, the analysis presented here can be generalized to describe more realistic situations by doing simple improvements like changing the particles' trajectories, which, in general, will not be geodesics. Yet, the idea that frame dragging could be measured in terrestrial laboratories is very appealing, particularly since it relies on the quantum nature of matter.

\section*{Acknowledgments}
We acknowledge getting valuable feedback from Abel Camacho. This research was funded by UNAM-DGAPA-PAPIIT Grant No. IA101818 and CONACyT through the graduate school scholarships.


\begin{thebibliography}{99}

\bibitem{Will}C. M. Will, \Journal{\em Liv. Rev. Rel.}{17}{4}{2014}.

\bibitem{COW}R. Colella, A. W. Overhauser, and S. A. Werner, \Journal{\PRL}{34}{1472}{1975}.

\bibitem{NoninertialCOW} U. Bonse and T. Wroblewski, \Journal{\PRL}{51}{1401}{1983}.

\bibitem{Bouchendira}R. Bouchendira {\it et al}, \Journal{\PRL}{106}{080801}{2011}.

\bibitem{Muller}P. Hamilton {\it et al}, \Journal{\em Science}{349}{(6250)849}{2015}.

\bibitem{Asenbaum}P. Asenbaum {\it et al}, \Journal{\PRL}{118}{183602}{2017}.

\bibitem{Wald} R. M. Wald, {\em General Relativity}, (The University Chicago Press, 1984).

\bibitem{WernerBook} H. Rauch and S. A. Werner, {\em Neutron Interferometry: Lessons in experimental quantum mechanics, wave-particle duality, and entanglement}, second edition, (Oxford University Press, 2015).

\bibitem{Hartle}J. B. Hartle, {\em Gravity: An introduction to Einstein’s General Relativity}, (Addison-Wesley San Francisco, 2003), chapters 8 and 9.

\bibitem{GPB}C. W. F. Everit {\it et al}, \Journal{\PRL}{106}{221101}{2011}.


\end{thebibliography}
\end{document}